\documentclass[pra,showpacs,preprintnumbers,amsmath,amssymb,tightenlines,twocolumn,superscriptaddress]{revtex4}

\usepackage{amsmath}
\usepackage{amsthm}
\usepackage{graphicx}
\usepackage{dcolumn}
\usepackage{bm}
\usepackage{epsfig}
\usepackage{stmaryrd}
\usepackage{color}

\newcommand{\bra}[1]{\langle\,{#1}\, |}
\newcommand{\ket}[1]{|\,{#1}\,\rangle}

\newcommand{\eref}[1]{Eq.~(\ref{#1})}

\newcommand{\cref}[1]{chapter~\ref{#1}}

\newcommand{\Cref}[1]{Chapter~\ref{#1}}

\usepackage{ulem}  
\begin{document}

\title{The equivalent emergence of time dependence in classical and  quantum  mechanics} 
\author{John S.\ Briggs}
\affiliation{Institute of Physics, University of Freiburg, Freiburg, Germany and\\
Department of Physics, Royal University of Phnom Penh, Cambodia}

\email{briggs@physik.uni-freiburg.de}
\begin{abstract}
Beginning with the principle that a closed mechanical composite system is timeless,  time can be defined by the regular changes in a suitable position coordinate (clock) in 
the observing part, when one part of the closed composite observes another part. Translating this scenario into both classical and quantum mechanics allows a transition to be made from a time-independent mechanics for the closed composite to a time-dependent description of the observed part alone. The use of Hamilton-Jacobi theory yields a very close parallel between the derivations in classical and quantum mechanics. The time-dependent equations, Hamilton-Jacobi or Schr\"odinger, appear as approximations since no observed system is truly closed. The quantum case has an additional feature in the condition that the observing environment  must become classical in order to define a real classical time variable. This condition leads to a removal of entanglement engendered by the interaction between the observed system and the observing environment. Comparison is made to the similar emergence of time in quantum gravity theory.
\end{abstract}
\maketitle

\section{Introduction}
In very many languages the same word or phrase is used to denote both spatial and temporal order. As the linguist Haspelmath \cite{Hasp} has observed ``But space and time seem to show a
peculiar relatedness that is perhaps not evident to a naive philosophical
observer: Human languages again and again express temporal and spatial
notions in a similar way" . Examples from the english language are the modern use of ``ahead of'' to imply
temporal precedence, or the older use of ``before'' to describe a spatial precedence.

The interchangeability of position and time is a consequence of the fact that the time concept is of human invention, arising from the attempt to quantify changes in position of observed objects by comparing with position changes of a standard object. A simple example is the motion of celestial bodies used to define time until quite recently. Early earth-bound objects include the sun-dial later superseded by the pendulum. All use position as time. To qualify to be categorised as what is known as a ``clock'' such a standard instrument must exhibit regularity in its position changes. The more regular, the more accurate the clock. Time is measured by position. The perfect clock would be a point particle having constant momentum. For practical reasons, real clocks are localised by enforcing circular or periodic position changes.

Consider, either in classical or quantum mechanics, a closed composite, denoted by $\mathcal{C}$ , consisting of two parts. One part, to be called the environment 
$\mathcal{E}$, observes
another part, to be called the system $\mathcal{S}$. It will be shown how a position coordinate of  $\mathcal{E}$ and its changes, can define a time parameter for changes in the mechanical coordinates of $\mathcal{S}$. 

The starting point, then, is a closed mechanical system, whether quantum or classical. Being closed it has a fixed total energy and therefore a time-independent Hamiltonian. In classical mechanics  the state is described by a structure in phase space subject to the constraint $(H(p,q) - E) = 0$ ; in quantum mechanics by a time-independent wave equation $(H - E)\Psi = 0$\,. Time is unnecessary in describing the dynamical state. Time can be introduced when one part of this composite observes the other that is, time parametrises changes in $\mathcal{S}$ when an observer (in the most general sense) constitutes the environment $\mathcal{E}$. 

To illustrate clearly that time emerges in exactly the same way in classical and quantum mechanics, a mathematical method will be used which allows the parallel to be traced most easily. This is the Hamilton-Jacobi (HJ) approach to classical mechanics. This close analogy led Schr\"odinger to his time-independent equation in the paper introducing wave mechanics \cite{sch1}. Incidentally Schr\"odinger \cite{sch2} had much greater difficulty with time dependence and one object of this paper is to give a clearer illustration of the path to a time-dependent equation.

It is interesting that the approach used here in non-relativistic classical and quantum mechanics is also a standard approach to the problem of time in quantum gravity \cite{Isham}
\cite{And}. 
The close similarity of the introduction of time into the TISE and the equivalent Wheeler-deWitt equation (WDE) of quantum gravity is outlined.

The plan of the paper is as follows.
First the time-independent HJ equation (TIHJE) of classical mechanics is derived for $\mathcal{C}$ by using a Jacobi time-independent variational principle for the path in phase space of a closed system. In particular generalised momenta will be defined in a time-independent way i.e without recourse to the standard definition in terms of derivatives with respect to velocities (as no time is defined, neither are velocities). In this way, momentum is interpreted as the propensity of objects to change position or, in the case of bound complexes, to change position and shape. This allows the Hamiltonian in the constraint $(H - E) = 0$ to be defined in terms of these momenta and so to derive the TIHJE in which momenta are given as derivatives
 of Hamilton's characteristic action function.
 
  In quantum mechanics essentially the same procedure is followed, defining the wavefunction as an exponential of an action function. 
 However, since this wavefunction is distributed over all space, following the recipe of the Cook books \cite{Cook},  the momentum must be interpreted as a momentum \emph{density} distribution. Then the time-independent Schr\"odinger equation (TISE) for $\mathcal{C}$ is derived by minimising the \emph{expectation value} of the constraint
 $(H - E) = 0$ over all space with respect to variations in the unknown wavefunction.
 
 These full equations TIHJE and TISE for $\mathcal{C}$ are then transformed to coupled equations by writing the total action function as a sum of a part for the observer $\mathcal{E}$ and a part for the observed $\mathcal{S}$ which depends parametrically on the coordinates of the observing $\mathcal{E}$. In both classical and quantum cases, the same approximations then lead to a partial de-coupling of the two parts and to
 a \emph{time-dependent} HJ equation (TDHJE) or a \emph{time-dependent} Schr\"odinger equation (TDSE) \emph{for the system} $\mathcal{S}$ \emph{alone}. In particular it is emphasised that the time appearing in quantum mechanics is always classical. 
 
 The paper closes with a commentary on the derivation of these time-dependent equations using different approximate quantum wavefunctions.  Comments are made also on the role of  entanglement in the transition from TISE for $\mathcal{C}$ to a TDSE for the observed part $\mathcal{S}$ and on how interacting environments lead to effective time-dependent potentials. The relevance of these results from non-relativistic quantum mechanics to the question of time in quantum gravity is discussed also.

 \section{Time-independent classical and quantum mechanics}
 
 In most text books on classical or quantum mechanics, the time-dependent equations, either TDHJE or TDSE, are considered the fundamental equations and the time-independent versions derived as the special case of a time-independent Hamiltonian. However, in this work space is taken as given but time is a relational quantity derived from space. Hence the time-independent equations involving space coordinates only, are viewed as fundamental.
 Hence these time-independent equations of classical mechanics, the TIHJE, and of quantum mechanics, the TISE,  are derived first, using time-independent variational principles.
 
 \subsection{Time independent  Classical Mechanics}			
Following and extending the treatment of Lanczos ÷\cite{lan}, ch.7,  the line element between two points is defined, 
\begin{equation} ds^{2} = f^{2}(q_{1}\ldots q_{n}) \sum_{ik} a_{ik} dq_{i}dq_{k} ,
\end{equation}
where the metric coefficients $a_{ik}$ in general are functions of the q's and the function $f$ is to be specified. The aim is to minimise the arc length
\begin{equation} 
\label{Eq:Wds}
 \mathit{W} = \int ds
\end{equation}
between the initial  point $(q^{0}_{1}\ldots q^{0}_{n})$ and the final point $(q_{1}\ldots q_{n}).$
Then $ds$ is written
\begin{equation} 
\label{Eq:ds}
 ds = f(\mathbf{q}) \sqrt{ \sum_{ik} a_{ik}dq_{i}dq_{k}}  = f(\mathbf{q})\sqrt{\mathbf{dqAdq}} 
\end{equation}
 For brevity in \eref{Eq:ds}  the coordinates $q_{i}$ have been written as the vector $\mathbf{q}$ (and correspondingly for the $\mathbf{dq}$) and the metric coefficients $a_{ik}$ as the matrix $\mathbf{A}$.

Now generalised  canonical momenta are introduced as,
\begin{equation}
\label{Eq:Genmom}
\begin{split}
 p_{j}& = \partial\Big(f(\mathbf{q})\:\sqrt{\mathbf{dqAdq}} \Big)/\partial{(dq_{j})}\\& = f(\mathbf{q}) \:{( \mathbf{dqAdq})}^{-1/2}\:{ (\mathbf{Adq})}_{j}.
 \end{split}
\end{equation}
Here the partial derivative $\partial (dq_i))$ has been introduced to quantify the explicit dependence of $W$ on the { \emph{changes}} $dq_i$ in the co-ordinates $q_i$. Such a definition was introduced by Barbour \cite{bar1}. Hence this definition gives momentum as a measure of the propensity of a body to change shape.  The usual definition of momentum in terms of velocities i.e. $\partial (dq_i/dt))$ requires the introduction of time, which is not defined for a closed system.
Then we can write the vector of the momenta as,

\begin{equation}
\label{Eq:defvecp}
 \mathbf{p} = f(\mathbf{q}) \:{( \mathbf{dqAdq})}^{-1/2}\:{ (\mathbf{Adq})}.
\end{equation}
This leads to a simple expression for the scalar product
\begin{equation} 
\mathbf{p\:dq} =  f(\mathbf{q})\sqrt{\mathbf{dqAdq}} 
\end{equation}
Hence, from \eref{Eq:Wds} and \eref{Eq:ds}, the integral to be minimised is the action,
\begin{equation}
\label{Eq:actionW}
\mathit{W}  = \int\sum_{j} p_{j}\:dq_{j}
\end{equation}
The minimisation is to be performed subject to the constraint, which is readily proved using \eref{Eq:defvecp},
\begin{equation}
 \mathbf{p\:B\:p} =  f^{2}(\mathbf{q}),
\end{equation}
where $\mathbf{B}$ is the inverse of the matrix  $\mathbf{A}$. The integral $\mathit{W}$ becomes the usual action integral when the choice 
\begin{equation}
f(\mathbf{q}) = \sqrt{ 2(E - V(\mathbf{q}))}
\end{equation}
is made. Then the constraint appears in the form of the conservation of total energy E and V is the potential energy, i.e.,
\begin{equation}
\label{Eq:pBp}
\frac{1}{2} \mathbf{p\:B\:p} + V(\mathbf{q}) = E.
\end{equation}
The function on the l.h.s. of this equation is called the Hamiltonian i.e.
\begin{equation}
\label{Eq:Hamdef}
H(\mathbf{p,q}) = \frac{1}{2} \mathbf{p\:B\:p} + V(\mathbf{q}).
\end{equation}
For this choice of constraint the metric elements $a_{ij}$ have physical dimensions of mass. However, it is simpler to choose the coordinates as mass-weighted i.e. 
$q_j = \sqrt{m_j} \tilde q_j$ when the elements $a_{ij}$ and  $b_{ij}$ are dimensionless. Then for the simple choice of metric $a_{ij} = \delta_{ij}$ we have the constraint
\begin{equation}
H = E = \frac{1}{2}\sum_{i}p_{i}^{2} + V.
\end{equation}
The action integral can be written in the simple form
\begin{equation}
\mathit{W} = \int\sqrt{\left(\mathbf{p\:B\:p}\right)( \mathbf{dqAdq})}.
\end{equation}

As Lanczos ÷\cite{lan} has emphasised, since the p's and q's are independent variables, in the form \eref{Eq:actionW} the end points of the integral must be varied. Indeed, rather than $\delta\mathit{W} = 0$, one has
\begin{equation}\delta\mathit{W} = \sum_{i}(p_{i}\delta q{i} - p_{i}^{0}\delta q_{i}^{0}).
 \end{equation}
With the choice \eref{Eq:pBp} leading to energy conservation $H = E$, the integral $\mathit{W}$ is identical to Hamilton's principal function. Then,
\begin{equation}
\delta{\mathit{W}} = \sum_{i}\left(\frac{\partial {\mathit{W}}}{\partial q_{i}}\delta q_{i} + \frac{\partial{ \mathit{W}}}{\partial q_{i}^{0}}\delta q_{i}^{0}\right).
 \end{equation}
Comparison of the two forms of  $\delta W$ gives,
\begin{equation}
\label{Eq:classmom}
 p_{i} = \frac{\partial {\mathit{W}}}{\partial q_{i}}
 \end{equation}
and
\begin{equation}
p_{i}^{0} =  - \frac{\partial{ \mathit{W}}}{\partial q_{i}^{0}}.
 \end{equation}
The $q_{i}^{0}$ can be taken to be $n$ constants of integration so that the constraint $H -E = 0$ becomes the time-independent Hamilton-Jacobi partial differential equation
(TIHJE),
 \begin{equation}
 \label{Eq:TIHJE}
H(q_{1} \ldots q_{n}, \frac{\partial {\mathit{W}}}{\partial q_{1}} \ldots \frac{\partial {\mathit{W}}}{\partial q_{n}}) - E = 0,
 \end{equation}
whose solution gives the relation between the $p$ and $q$ variables. The dynamics are timeless as is the definition of generalised momentum \eref{Eq:Genmom}.

\subsection{Time independent Quantum Mechanics}
One notes that the classical TIHJE is the constraint $H = E$ for a closed system, where the path variation has led to the identification of momentum with the derivative of the action function $W$. Now exactly the same results will be employed to derive the TISE.
 This method is actually that of Schr\"{o}dinger's 1926 papers ÷\cite{sch1,sch2},  although little used in standard quantum mechanics books today (see, however, Cook \cite{Cook}).  Schr\"{o}dinger began with the TIHJE and Hamilton's characteristic function derived above for classical mechanics. He postulated a wave function $\Psi$ such that 
\begin{equation}
\mathit{W} = k \ln{\Psi}\;\; \mathrm{or}\;\; \Psi = \exp\left( \frac{1}{k}\mathit{W}\right ),
\end{equation}
where $k$ is a real constant. In fact in his first derivation Schr\"{o}dinger did not admit  complex wavefunctions. With the benefit of hindsight one takes the functions
\begin{equation}
\mathit{W} = -i\hbar \ln{\Psi}\;\; \mathrm{or}\;\; \Psi = \exp\left( \frac{i}{\hbar}\mathit{W}\right ),
\end{equation}
where $W$ is in general complex. Note that this means that the $i$ in the exponential could be dropped, it is retained by convention. This logarithmic relation is motivated by the observation that for many particles the classical action is additive, whereas the wavefunction is multiplicative. The use of a complex action in the quantum case, leading to a complex wavefunction, whereas the classical action is real, has been the subject of some discussion \cite{Cramer,And1,bar2} and will be justified below in section V.

The quantum momenta are defined exactly as in the classical case \eref{Eq:classmom},
\begin{equation}
\label{Eq:quantump}
p_{i} = \frac{\partial{\mathit{W}}}{\partial{q_{i}}} = -i\hbar \left(\frac{1}{\Psi}\frac{\partial{\Psi}}{\partial{q_{i}}}\right).
\end{equation}
Recognising that the kinetic energy must be real, a direct substitution of \eref{Eq:quantump} in \eref{Eq:TIHJE}) gives rise to the equation,
\begin{equation}
\label{Eq:quantum H}
\frac{1}{2}\sum_{i}\left|-i\hbar \left(\frac{1}{\Psi}\frac{\partial{\Psi}}{\partial{q_{i}}}\right)\right|^{2} + V - E = 0.
\end{equation}
Such a non-linear differential equation is not appropriate to describe matter waves. Rather, as Cook has emphasised \cite{Cook},  the wavefunction $\Psi$ represents a distribution over all space and the quantum $|p_i|^2$ is a momentum density. Then the optimum form of $\Psi$ is obtained from minimising the mean value of the H-J  constraint equation over all space, i.e. one demands
 \begin{equation}
\delta\int\left|\Psi\right|^{2}\left(H(\mathbf{q},\frac{\partial{\mathit{W}}}{\partial{\mathbf{q}}}) -E\right) d\mathbf{q} = 0,
\end{equation}
with the momenta to be substituted from \eref{Eq:quantump}. This is then the Euler variational problem,
 \begin{equation}
 \begin{split}
\delta\int\ \left(\sum_{i}\frac{\hbar^2}{2}\left(\frac{\partial{\Psi^{\ast}}}{\partial{q_{i}}}\frac{\partial{\Psi}}{\partial{q_{i}}}\right) + \Psi^{\ast}\left(V - E\right)\Psi\right)& dq_{1}\ldots \;dq_{n}\\& = 0,
\end{split}
\end{equation}
with $\Psi = \Psi(q_{1}\ldots q_{n})$. This variational principle leads to the differential equation,
\begin{equation}
 \begin{split}
&-\frac{\hbar^2}{2}\sum_{i} \left(\frac{d}{dq_{i}}\frac{\partial{\Psi}}{\partial{q_{i}}}\right) + (V -E)\Psi\\& = -\frac{\hbar^2}{2} \sum_{i} \frac{\partial^{2}{\Psi}}{\partial{q_{i}^{2}}} +  (V -E)\Psi = 0,
\end{split}
\end{equation}
i.e. to the TISE for the composite of environment and system. Inclusion of the mass scaling would give the appropriate mass factors.  Note also
that  formally one can write \eref{Eq:quantump} as the ``eigenvalue" equation
\begin{equation}
p_{i}\Psi = -i\hbar \left(\frac{\partial{\Psi}}{\partial{q_{i}}}\right)
\end{equation}
which illustrates the origin of this differential form. Again, as in classical mechanics, for the quantum dynamics of the composite $\mathcal{C}$ at fixed energy $E$, no time is defined.

\section{The emergence of time in classical mechanics}
 Time emerges when the composite is separated into system and environment and one or more of the environment variables are taken as the clock variables. The environment coordinates will be denoted by,
\begin{equation} \mathbf{R} = R_{1} \ldots R_{l} \equiv q_{1} \ldots q_{l}
 \end{equation}
and the system coordinates by,
\begin{equation}
\mathbf{x} = x_{1} \ldots x_{n-l} \equiv q_{l+1} \ldots q_{n}.
 \end{equation}
without loss of generality, the total action $W$ can be written as a sum of system and environment actions in the form,
\begin{equation}
\label{Eq:adiabatW}
W(\mathbf{x,R}) = W_{\varepsilon}(\mathbf{R}) + W_{\mathcal{S}}(\mathbf{x,R}),
 \end{equation}
i.e., the system depends parametrically on the state of the environment.  Although not necessary, it is by far simpler if initially only a single variable $x$ for the system and
 $R$ for the environment, are considered.
 Then, with $H = T + V$, the total potential energy can be decomposed  into the potential energy  $V_{\varepsilon}$ of the environment alone, the potential energy $V_{\mathcal{S}}$ of the system alone and the potential energy $V_I$ of the necessary interaction of the environment with the system, i.e. 
the total Hamiltonian is written
\begin{equation}
\label{Eq:classHE}
H = T + V = T + V_{\varepsilon} + V_{S} + V_I = E.
 \end{equation}
 The total kinetic energy $T$  given by, 
\begin{equation}
T = \frac{1}{2}\sum_{i=1}^{n} \left(\frac{\partial {W}}{\partial q_{i}}\right)^{2}.
 \end{equation}
which, for just two degrees of freedom, reduces to
\begin{equation}
\label{Eq:fullHJ}
\begin{split}
T = \frac{1}{2}\Bigg[ \left(\frac{\partial {W_{\varepsilon}}}{\partial R}\right)^{2}
& + 2 \left(\frac{\partial {W_{\varepsilon}}}{\partial R}\frac{\partial {W_{\mathcal{S}}}}{\partial R}\right)\\& +  \left(\frac{\partial {W_{\mathcal{S}}}}{\partial R}\right)^{2} +  \left(\frac{\partial {W_{\mathcal{S}}}}{\partial x}\right)^{2}\Bigg].
 \end{split}
 \end{equation}

 The constraint equation can be split into two parts as follows. First, to make the physical dimensions clear, the mass scaling of coordinates will be abandoned and the mass $M$ of $\mathcal{E}$ and $m$ of $\mathcal{S}$ indicated explicitly. 
Then, all the terms not dependent upon $x$ are grouped on the r.h.s of the equation to give
 \begin{equation}
\begin{split}
& \frac{1}{2m} \left(\frac{\partial {W_{\mathcal{S}}}}{\partial x}\right)^{2} + V_{\mathcal{S}} + V_I\\ &+ \frac{1}{M}\left(\frac{\partial {W_{\varepsilon}}}{\partial R}\frac{\partial {W_{\mathcal{S}}}}{\partial R}\right) + \frac{1}{2M}\left(\frac{\partial {W_{\mathcal{S}}}}{\partial R}\right)^{2}\\& = - \left( \frac{1}{2M} \left(\frac{\partial {W_{\varepsilon}}}{\partial R}\right)^{2} +  V_{\varepsilon}(R)\right)+ E
 \end{split}
\end{equation}
which can be written
 \begin{equation}
 \label{Eq:splitclass}
\begin{split}
& H_{\mathcal{S}} + V_I\\ &+ \frac{1}{M}\left(\frac{\partial {W_{\varepsilon}}}{\partial R}\frac{\partial {W_{\mathcal{S}}}}{\partial R}\right) + \frac{1}{2M}\left(\frac{\partial {W_{\mathcal{S}}}}{\partial R}\right)^{2}\\& = - H_{\varepsilon} + E.
 \end{split}
\end{equation}
Now both sides of the equation can be put equal to a function of $R$ only, call it $U_{\mathcal{S}}(R)$. This gives an equation for  changes of $\mathcal{E}$ dependent
on the state of the system,
\begin{equation}
\label{Eq:HJenviron}
H_{\varepsilon} \equiv  \frac{1}{2M} \left(\frac{\partial {W_{\varepsilon}}}{\partial R}\right)^{2} +  V_{\varepsilon}(R)  = E - U_{\mathcal{S}}(R).
 \end{equation}
and correspondingly the system change depends upon the state of the environment,
\begin{equation}
\begin{split}
\label{Eq:HJsystem}
H_{\mathcal{S}} + V_I +  \frac{1}{M} \left(\frac{\partial {W_{\varepsilon}}}{\partial R}\frac{\partial {W_{\mathcal{S}}}}{\partial R}\right)& +  \frac{1}{2M}\left(\frac{\partial {W_{\mathcal{S}}}}{\partial R}\right)^{2} \\&= U_{\mathcal{S}}(R).
 \end{split}
\end{equation}
These two equations, which are still exact, indicate the coupling between environment and system. Energy can be transferred between the two parts, only the sum $E$ is conserved.
 
The effective energy for both $\mathcal{E}$ and $\mathcal{S}$ varies with $R$ according to the magnitude of $U_{\mathcal{S}}(R)$. However, for the environment to function as a clock it must have a regular variation in position not dependent upon the state of the system. This requires that $U_{\mathcal{S}}(R)$  in \eref{Eq:HJenviron} be neglected or replaced by a constant average value. In turn this implies that $V_I$, although finite, must be small compared to the clock energy.  Then one has a fixed clock Hamiltonian
\begin{equation}
\label{Eq:HJenviron1}
 \frac{1}{2M} \left(\frac{\partial {W_{\varepsilon}}}{\partial R}\right)^{2} +  V_{\varepsilon}(R) = E_c
 \end{equation}
where $E_c$ is a fixed energy of the clock. In this approximation energy conservation for the composite $\mathcal{C}$ as a whole is abandoned.

Now the condition of a minimum invasion of the system by the environment (clock) is imposed in that the dependence of the system on the environment variables through $\partial W_{\mathcal{S}}/\partial R$  will be assumed small. Then the quadratic term in $\partial W_{\mathcal{S}}/\partial R$ in  \eref{Eq:HJsystem} can be neglected.
 The validity and consequences of this are discussed further below.
 At this stage a transition from a position $R$ dependence to a time dependence for the system can be made.

The clock time is introduced by defining the environment parameter,
\begin{equation}
\label{Eq:deftime1}
t = M\int^{R}\frac{dR'}{p(R')},
 \end{equation}
 with $p(R) = \partial {W_{\varepsilon}}/\partial R =  [2M(E_c - V_{\varepsilon}(R)]^{1/2}$.
Then $\frac{1}{M}\partial {W_{\varepsilon}}/{\partial R} = dR/dt$ and the cross term in  \eref{Eq:HJsystem} can be written,
\begin{equation}
\label{Eq:classdt}
 \frac{1}{M}\left(\frac{\partial {W_{\varepsilon}}}{\partial R}\frac{\partial {W_{\mathcal{S}}}}{\partial R}\right)  =   \left(\frac{dR}{dt} \frac{\partial {W_{\mathcal{S}}}}{\partial R}\right) =  \frac{\partial {W_{\mathcal{S}}}}{\partial t}.
 \end{equation}
This changes the parametric $R$ dependence of $\mathcal{S}$ into a parametric $t$ dependence and \eref{Eq:HJsystem}  becomes,
\begin{equation}
\label{Eq:TDHJE1}
  H_{\mathcal{S}}(x)   + V_I (x,t)+  \frac{\partial {W_{\mathcal{S}}}}{\partial t}  = U_{\mathcal{S}}(t).
\end{equation}

 The energy $U_{\mathcal{S}}$ does not depend upon $x$ and can be transformed away by the substitution $\mathrm{S}(x,t) = W_{\mathcal{S}}(x,t) - \int^t U_{\mathcal{S}}(t') dt'$  (here the usual notation $\mathrm{S}$ for a time-dependent action function has been introduced).  This gives the time-dependent Hamilton-Jacobi equation
 (TDHJE) for the observed system,
\begin{equation}
\label{Eq:TDHJE2}
 \frac{1}{2m} \left(\frac{\partial {\mathrm{S}(x,t)}}{\partial x}\right)^{2} + V_{\mathcal{S}}(x) + V_I(x,t) +  \frac{\partial {\mathrm{S}(x,t)}}{\partial t} = 0
\end{equation}
or, 
\begin{equation}
\label{Eq:TDHJE3}
H_{S}\left(x,\frac{\partial {\mathrm{S}(x,t)}}{\partial{x}}\right) + V_I(x,t) +  \frac{\partial {\mathrm{S}}}{\partial t} = 0.
\end{equation}
Note again that the parametric $R$ dependence has now been replaced by the parametric time dependence. To monitor the 'true' time dependence of  system $\mathcal{S}$ the interaction potential $V_I$ has to be negligibly small. 

The foregoing analysis shows explicitly how the parameter of time, arising from a spatial correlation of the \emph {position} coordinates of system and clock, enters into classical mechanics. The usual Hamilton and Newton time-dependent equations of the system dynamics (but where time is not assumed, as Newton and Hamilton did, simply to exist, rather it arises from comparison with the position coordinate of a material clock) follow directly from the TDHJE of \eref{Eq:TDHJE3}.
 The key element in the derivation of the TDHJE for the system is the neglect of the kinetic energy term 
$(\partial {W_{\mathcal{S}}}/\partial R)^2$. If  this term cannot be neglected one obtains corrections to Hamilton's and Newton's equations of motion. This point is discussed further in section V C.

\section{The emergence of classical time in quantum mechanics}

To summarise the results of the previous section. Two approximations have led from the TIHJE for the composite $\mathcal{C}$ to the TDHJE for the system $\mathcal{S}$ alone. One is that the term 
$(\partial W_{\mathcal{S}}/\partial R)^2$ term can be neglected. In addition, in order to function as a clock, the back-reaction of the interaction potential has been
neglected in obtaining a fixed-energy TIHJE for $\mathcal{E}$ alone. Then the total TIHJE for $\mathcal{C}$ separates into a TIHJE for $\mathcal{E}$ and a TDHJE for the observed system 
$\mathcal{S}$. The environment kinetic energy provides the time variable for the system via the $(\frac{\partial W_\varepsilon}{\partial R})/M$ velocity term. Now it will be shown that exactly the same approximations in the quantum mechanics case allow the derivation of  a TDSE for $\mathcal{S}$ alone from the TISE for $\mathcal{C}$.

The starting point is  the full TISE in the form
\begin{equation}
\label{Eq:full2dTISE}
\begin{split}
& -\frac{\hbar^2}{2M}\left(\frac{\partial^2\Psi}{\partial R^2}\right)  -\frac{\hbar^2}{2m}\left(\frac{\partial^2\Psi}{\partial x^2}\right)\\&
 + [V_{\varepsilon}(R) +  V_{\mathcal{S}} + V_I(x,R) - E] \Psi(x,R) = 0
 \end{split}
\end{equation}
which is the analogue of the classical \eref{Eq:classHE}.
Now one takes an action function exactly of the form
 \eref{Eq:adiabatW}
\begin{equation}
W(x,R) = W_{\varepsilon}(R) + W_{\mathcal{S}}(x,R),
 \end{equation}
 of the classical case, 
but to conform with quantum mechanics these action functions may be complex. Then the total wavefunction is written
\begin{equation}
\label{Eq:prodwf}
\begin{split}
\Psi(x,R)& = \exp{\left(\frac{i}{\hbar}W(x,R)\right)}\\& = \exp{\left(\frac{i}{\hbar}[W_{\varepsilon}(R) + W_{\mathcal{S}}(x,R)]\right)} \\&\equiv \chi(R)~\psi(x,R)
\end{split}
\end{equation}
This product form is of adiabatic type but is exact at the moment \cite{Ced}.

Substitution in the TISE gives the equation
\begin{equation}
\label{Eq:TISEprod}
\begin{split}
\chi \left(H_\mathcal{S} + V_I  - \frac{\hbar^2}{M}\frac{1}{\chi}\frac{\partial \chi}{\partial R}\frac{\partial}{\partial R} - \frac{\hbar^2}{2M} \frac{\partial^2}{\partial R^2}\right)\psi&\\ 
= - \psi \left( H_\mathcal{E} - E\right)\chi.
\end{split}
\end{equation}
which is the direct term-by-term analogue of the classical \eref{Eq:splitclass}.
Integration over the space of $x$ only, denoted by round brackets, gives the defining equation for $\chi$
\begin{equation}
\label{Eq:chidef}
\begin{split}
\left(- \frac{\hbar^2}{2M} \frac{\partial^2}{\partial R^2} + V_\mathcal{E} \right )\chi
= (E -  U_\mathcal{S}(R))\chi
\end{split}
\end{equation}
with the definition
\begin{equation}
\label{Eq:Uquant}
\begin{split}
&(\psi|H_{\mathcal{S}}(x) + V_{I}(x,R) \\& - \frac{\hbar^2}{M}\frac{1}{\chi} \frac{\partial \chi}{\partial R}\frac{\partial}{\partial R} - \frac{\hbar^2}{2M} \frac{\partial^2}{\partial R^2}|\psi) \equiv U_\mathcal{S}(R).
\end{split}
\end{equation}
These last two equations are the transcription of the classical equations \eref{Eq:HJenviron} and \eref{Eq:HJsystem} as may be seen by inspection.

The equation for the wavefunction $\psi$ is then
\begin{equation}
\label{Eq:psidef}
\begin{split}
\Bigg(H_{\mathcal{S}} + V_I(x,R) - U_\mathcal{S}(R) - \frac{\hbar^2}{2M} \frac{\partial^2}{\partial R^2}\\ -  \frac{\hbar^2}{M}\frac{1}{\chi}\frac{\partial \chi}{\partial R}\frac{\partial }{\partial R}\Bigg)\psi = 0.
\end{split}
\end{equation}
 Note that in the quantum equations the kinetic energy terms appear as
 second derivatives, in the classical equation as a quadratic term in first derivatives. This is a standard result of the calculus of variations and is the difference between deterministic
classical and probabilistic quantum mechanics.

Although exact, the two equations (\ref{Eq:chidef}) and (\ref{Eq:psidef}) are not easy to solve since they are strongly coupled in that $\psi$ appears in \eref{Eq:chidef} and
$\chi$ in  \eref{Eq:psidef}. To allow the environment to function as a clock, exactly the same approximations as in the classical case are necessary.
 Firstly, the back-coupling of the system on the clock must be neglected by ignoring $U_\mathcal{S}(R)$ or putting it equal to a constant in \eref{Eq:chidef}. 
In the classical case this suffices to define time as in \eref{Eq:deftime1}. In the quantum case, since time is a classical quantity, the further step of going to a \emph{classical} limit for the environment action $W_{\varepsilon}(R)$ is necessary. This one sees by putting 
\begin{equation}
\chi = \exp{\left(\frac{i}{\hbar}W_{\varepsilon}\right)} 
\end{equation}
 in \eref{Eq:chidef}, now with fixed $E_c \approx E - U_\mathcal{S}(R)$.
This gives
\begin{equation}
\label{Eq:qenviron}
 \frac{1}{2M} \left(\frac{\partial {W_{\varepsilon}}}{\partial R}\right)^{2} - i  \frac{\hbar}{2M}\frac{\partial^2 {W_{\varepsilon}}}{\partial R^2} +  V_{\varepsilon}(R) = E_c
 \end{equation}
The difference compared to the classical \eref{Eq:HJenviron1} is the second derivative in $R$ arising from the transition to quantum mechanics. Then this term must be neglected
to obtain the WKB approximation with $W_{\varepsilon}$ the real classical action leading to a real classical time.

Following the transition of the environment to classical mechanics, in \eref{Eq:psidef} the cross term becomes
\begin{equation}
\begin{split}
\frac{\hbar^2}{M}\frac{1}{\chi}\frac{\partial \chi}{\partial R}\frac{\partial\psi }{\partial R}& = i\hbar\frac{1}{M}\frac{\partial W_{\varepsilon}}{\partial R}\frac{\partial\psi }{\partial R}
\\&
=~ i\hbar\frac{\partial\psi}{\partial t}.
\end{split}
\end{equation}
to give a time derivative exactly as in the classical \eref{Eq:classdt}.
Secondly, the ``crucial approximation" to neglect $\left(\partial {W_{\mathcal{S}}}/\partial R\right)^{2}$ in the classical case, translates into the neglect of the term $ \partial^2\psi/\partial R^2$ in the quantum  \eref{Eq:psidef}. 

With the quantum $R$ dependence replaced by a classical $t(R)$ dependence then \eref{Eq:psidef} reduces to
\begin{equation}
\left(H_{\mathcal{S}} + V_I(x,t) - U_\mathcal{S}(t) -  i\hbar\frac{\partial}{\partial t} \right)\psi = 0.
\end{equation}
As in the classical case the purely time-dependent potential $U_\mathcal{S}(t)$ can be removed, here by a phase transformation, to yield the TDSE
\begin{equation}
\label{Eq:TDSE1}
\left(H_{\mathcal{S}} + V_I(x,t) -  i\hbar\frac{\partial}{\partial t} \right)\psi = 0.
\end{equation}
for the quantum system alone. In full this TDSE reads
\begin{equation}
\label{Eq:TDSE2}
- \frac{\hbar^2}{2m} \frac{\partial^2\psi(x,t)}{\partial x^2} + \left(V_{\mathcal{S}} + V_I(x,t) -  i\hbar\frac{\partial}{\partial t} \right)\psi(x,t) = 0
\end{equation}
which is to be compared to the equivalent TDHJE of \eref{Eq:TDHJE2}. This demonstrates that exactly the same physical approximations lead to time dependence of a system observed by an environment in both classical and quantum mechanics.

\section{Commentary}

To summarise, a closed composite at fixed total energy $E$, comprised of an environment and a system interacting via a potential $V_I$ has been considered.
Beginning with the classical constraint $(H - E) = 0$, a TIHJE for the composite has been derived in deterministic classical mechanics. This involves the action functions of $\mathcal{E}$ and
$\mathcal{S}$. In probabilistic quantum mechanics an analogue equation has been derived from the constraint $(\bra{\Psi}H\ket{\Psi} - E) = 0$ where $H$ is interpreted as a Hamiltonian density.
Then a variational principle for the form of this unknown wavefunction leads to the TISE for the composite. 

An Ansatz for the total action as a sum of two parts, with the observed system part dependent parametrically upon the environment coordinate, is then made as in \eref{Eq:adiabatW}. In the classical case subject to the validity of two approximations, this leads to two equations, a TIHJE for $\mathcal{E}$ and a TDHJE for $\mathcal{S}$. In an exactly analogous way, the same two approximations lead to a TISE  for $\mathcal{E}$ and a TDSE for $\mathcal{S}$. In the quantum case, additionally the environment must be taken in its classical TIHJE limit to define a real time variable. In both cases it is the same part of the kinetic energy of the environment, the changes in position of the ``clock", that provide the classical time variable for the system $\mathcal{S}$.

The method of derivation has been tailored specifically to highlight the extremely close similarity of the emergence of time in classical and quantum mechanics.
However, the quantum case allows much more flexibility in that the linear TISE, in contrast to the non-linear TIHJE,  admits of sums of products as a solution as is discussed next.

\subsection{Superposition wavefunctions and entanglement}

As mentioned, the equations (\ref{Eq:chidef}) and (\ref{Eq:psidef}) for the two factors of the exact single product wavefunction are strongly coupled. A different set of equations are obtained if an, in principle infinite, sum is made over product wavefunctions, one set of which is of a \emph{pre-specified} type. The most common example from molecular physics is the Born-Oppenheimer (BO) expansion,
\begin{equation}
\label{Eq:BOexp}
\Psi(x,R) = \sum_n \chi_n(R)\psi_n(x,R)
\end{equation}
where now the $\psi_n$ are not to be determined but are fixed at the outset as the various eigenstates of the equation
\begin{equation}
(H_{\mathcal{S}}(x) + V_I(x,R) )\ket{\psi_n} = U^{BO}_n(R)\ket{\psi_n}
\end{equation}
for each fixed value of $R$. This is also called the adiabatic expansion and its use in deriving the TDSE is described in \cite{br12} .

Several authors \cite{Ced}, \cite{Arce} have shown that, although ostensibly the same as a single-channel BO form, the single product  wavefunction of
 \eref{Eq:prodwf} is an \emph{exact} representation of the  total wavefunction (as is obvious if one formally sets $\psi = \Psi/\chi$). From the single product form one could also infer that this represents a non-entangled state. However that this is a fully-entangled wavefunction is seen readily if one expands
\begin{equation}
\label{Eq:expphi}
\psi(x,R) = \sum_n c_n(R)\phi_n(x)
\end{equation}
so that 
\begin{equation}
\label{Eq:entang2}
\begin{split}
\Psi(x,R) = \chi(R) \sum_n c_n(R)\phi_n(x)& =  \sum_n \chi(R)c_n(R)\phi_n(x)
\\& \equiv  \sum_n \kappa_n(R)\phi_n(x).
\end{split}
\end{equation}

Clearly such a form is an exact representation if the sum covers the whole Hilbert space of both parts. Introducing this form into the full TISE $(H - E)\Psi = 0$
 gives the set of  coupled equations
\begin{equation}
\label{Eq:close coupled}
[H_{\varepsilon} - E] \kappa_m(R) = - \sum_n (\phi_m| H_{\mathcal{S}} + V_I |\phi_n) \kappa_n(R)
\end{equation}
where round brackets indicate integration over $x$ only.

To derive the system TDSE  in the case of the entangled state expansion \eref{Eq:entang2}, one writes explicitly
\begin{equation}
\begin{split}
\kappa_m(R) = c_m(R) \exp{\left(\frac{i}{\hbar} W(R)\right)}.
\end{split}
\end{equation}
Here $W$ is real and as yet unspecified but the $b_m$ are complex functions.
Substitution in \eref{Eq:close coupled} gives the equivalent equations
\begin{equation}
\begin{split}
 &\left[ \frac{1}{2M}\left(\frac{\partial W}{\partial R}\right)^{2}  + \frac{\hbar
^{2}}{2M}\frac{\partial ^{2}}{\partial R^{2}} + V_{\mathcal{E}}(R) - E\right]  c_{m}\\& 
 + \left[\frac{i\hbar }{2M}c_{m}\frac{\partial ^{2}W}{\partial R^{2}} + \frac{i\hbar
}{M}\frac{\partial c_{m}}{\partial R}\frac{\partial W}{\partial R}\right]  \\
 & +  \sum_n( \phi _{m}\left\vert H_{\mathcal{S}} +V_{I}\right\vert \phi _{n}) 
c_{n}(R)=0.
\end{split}
\end{equation}

As before, to derive the TDSE, the second derivatives w.r.t $R$ must be neglected. This gives the simpler equations,
\begin{equation}
\begin{split}
 & \left[ \frac{1}{2M}\left( \frac{\partial W}{\partial R}\right) ^{2} + V_{\mathcal{E}}(R) - E\right]  c_{m}(R) = 
- \left[\frac{i\hbar
}{M}\frac{\partial c_{m}}{\partial R}\frac{\partial W}{\partial R}\right]  \\& 
 + \sum_{n}( \phi _{m}\left\vert H_{\mathcal{S}} +V_{I}\right\vert \phi _{n}) 
c_{n}{ (}R{ )}. 
\end{split}
\end{equation}
Here the terms on the r.h.s. play the role of the potential $U_{\mathcal{S}}(R)$ in the product expansion. This form illustrates  that the states $c_m(R)$ of the 
environment are dependent upon \emph{all} possible states of excitation of the system. In this form one sees very clearly that \emph{choosing} $W$
 to be the  classical action causes the l.h.s. of the above equation to become zero. If for simplicity, although not necessary, one chooses the $\phi_n$ to diagonalise $H_{\mathcal{S}}$ with eigenenergies $\epsilon_n$ one has the simpler coupled equations
\begin{equation}
\begin{split}
&\epsilon_m c_m(R) - \frac{i\hbar}{M}\frac{\partial c_{m}}{\partial R}\frac{\partial W}{\partial R} \\&  + \sum_{n}( \phi _{m}\left\vert V_{I}\right\vert \phi _{n}) 
c_{n}{ (}R{ )} =0.
\end{split}
\end{equation}
Proceeding exactly as before to define time through $R(t)$ and introducing the phase transformation
\begin{equation}
c_m(t) = a_m(t) \exp{\left(\frac{i}{\hbar}\epsilon_mt\right)}
\end{equation}
leads to the coupled equations, 
\begin{equation}
\begin{split}
\label{Eq:coupltime}
& i\hbar\frac{\partial a_m }{\partial t}\\& =\sum_{n}( \phi _{m}\left\vert V_{I}(t)\right\vert \phi _{n}) 
a_{n}{ (}t{ )} \exp{\left(\frac{i}{\hbar}(\epsilon_m - \epsilon_n)t\right)} .
\end{split}
\end{equation}
Now the \emph{environment} wavefunction amplitudes $c_m(R)$ have been transformed to the occupation amplitudes $a_m(t)$ of the \emph{system} eigenstates.
This set of equations is equivalent to the TDSE for the system, as is readily seen by substituting the expansion
\begin{equation}
\psi(x,t) = \sum_n a_n(t) \phi_n(x) \exp{\left(\frac{i}{\hbar}\epsilon_n t\right)}\end{equation}
in the TDSE \eref{Eq:TDSE1} and projection on $(\phi_m|$.

This derivation illustrates concisely how the entanglement between environment and system expressed through the dependence of environment  spatial wavefunctions upon the occupation of a given
system state $n$, is transformed into a coupling between time-dependent complex system amplitudes. Also it makes clear that the environment must be describable by a classical action in order to obtain a real time variable.

Entanglement is of course entirely dependent upon the interaction $V_I$ between $\mathcal{E}$ and $\mathcal{S}$. Were this zero, the composite is separable and there
is no entanglement.
In the case of the product form  \eref{Eq:prodwf} the entanglement  is not so transparent since it arises not only from the interaction potential but also from
the derivatives with respect to $R$. In particular the term
\begin{equation}
\left(\frac{\hbar^2}{M}\frac{1}{\chi}\frac{\partial \chi}{\partial R}\frac{\partial }{\partial R}\right)\psi 
\end{equation}
 in \eref{Eq:psidef} 
which ultimately becomes the $\partial/\partial t$ in the classical limit represents quantum entanglement of environment and system, as is seen from the expansion
\eref{Eq:expphi},
\begin{equation}
\begin{split}
\frac{\hbar^2}{M}\frac{1}{\chi}\frac{\partial \chi}{\partial R}\frac{\partial\psi}{\partial R} =  \frac{\hbar^2}{M}\frac{1}{\chi}\frac{\partial \chi}{\partial R}\sum_n\frac{\partial c_n(R)}{\partial R}\phi_n(x)
\end{split}
\end{equation}

Although the environment wavefunction $\chi$ is retained at the semi-classical level, to give the $\partial/\partial t$ parametric derivative from the expression above, one sees that the only elements appearing in the TDSE for the quantum system are this derivative and
the classical time $t(R)$. Hence, there is no remnant of quantum entanglement between environment and system in this approximation.
There is of course correlation through interaction, directly by the environment on the quantum system via the operator $V_I$ and a back reaction of the quantum system on the environment via the potential $U_\mathcal{S}(R)$. 

 Since the coupling element 
\begin{equation}
 \frac{1}{M}\left(\frac{\partial {W_{\varepsilon}}}{\partial R}\frac{\partial {W_{\mathcal{S}}}}{\partial R}\right)
\end{equation}
 leading to time dependence $\partial/\partial t$ in the classical equation is of  the same form 
\begin{equation}
\frac{\hbar^2}{M}\frac{1}{\chi}\frac{\partial \chi}{\partial R}\frac{\partial\psi}{\partial R} = i\frac{\hbar}{M}\frac{\partial {W_{\varepsilon}}}{\partial R}\frac{\partial {W_{\mathcal{S}}}}{\partial R}
\end{equation}
as that in the quantum case, this provokes the question as to the interpretation of the quantum entanglement and its classical counterpart.  However, although \eref{Eq:adiabatW} leads to the product entangled wavefunction and so could be interpreted as ``classical'' entanglement, there is no separation of the classical action corresponding to the \emph{sum} over products $\sum_n\kappa_n(R)\phi_n(x)$. It is more correct to say that the nature of the  \emph{correlation} between $\mathcal{E}$ and $\mathcal{S}$ is the same in classical and quantum mechanics.

\subsection{Interacting environments}
In the development of time dependent equations the case of a clock as environment has been considered. The interaction $V_I$ has been taken as negligibly small to minimise the back-coupling on the clock. However, most books tacitly assume that $V_I$ is identically zero. Then the time appearing is presumably some absolute Newtonian time.
That is, when $V_I$ is put to zero and the neglect of $ \left(\frac{\partial {W_{\mathcal{S}}}}{\partial R}\right)^{2}$ is taken as exact, the system Hamiltonian is time independent in both classical and quantum mechanics. In this approximation the classical TDHJE  \eref{Eq:TDHJE1} assumes the form
\begin{equation}
\label{Eq:TDHJE4}
 H_{\mathcal{S}}\left(x, \frac{\partial {\mathrm{S}}}{\partial x}\right) + \frac{\partial {\mathrm{S}}}{\partial t} = 0
\end{equation}
However, since $W_{\mathcal{S}}$ is now time independent one can put $S = W_{\mathcal{S}} - E_{\mathcal{S}}t$ to give the TIHJE for the system
\begin{equation}
\label{Eq:TIHJsystem}
 H_{\mathcal{S}}\left(x,\frac{\partial{W_{\mathcal{S}}}}{\partial x}\right) = E_{\mathcal{S}}.
\end{equation}
 Now of course the system is considered closed and no time is necessary. Nevertheless this TDHJE \eref{Eq:TDHJE4} is usually considered more ``fundamental'' than the TIHJE
 \eref{Eq:TIHJsystem}.
 
 An analogous situation is encountered in quantum mechanics. If $V_I$ is taken to be identically zero in \eref{Eq:TDSE2} then the Hamiltonian is time-independent and
 \eref{Eq:TDSE1} becomes the TDSE
\begin{equation}
\label{Eq:TDSE3}
H_{\mathcal{S}}(x)\psi(x,t) - i\hbar\frac{\partial\psi(x,t)}{\partial t} = 0.
\end{equation}
Again, however, the quantum system is now closed so that the simple phase transformation
\begin{equation}
\label{Eq;timephase}
\psi(x,t) = \phi(x) \exp{\left(-\frac{i}{\hbar}E_{\mathcal{S}}t\right)}
\end{equation}
 leads to the TISE
 \begin{equation}
 H_{\mathcal{S}}(x)\phi = E_{\mathcal{S}}\phi.
\end{equation}
Interestingly, even very prominent physicists \cite{Fine} have interpreted \eref{Eq;timephase} as implying that the eigenfunction of a time-independent quantum system oscillates in time - a rather remarkable behaviour for a closed system without time !

Most authors simply generalise the TDHJE and TDSE above for time-independent Hamiltonians, where the time is spurious, to time-dependent Hamiltonians, where, unlike for the clock, the interaction $V_I(x,t)$ is not negligible but drives transitions in the system.

 In the formalism developed here, this time dependence of the Hamiltonian appears naturally.  The environment is taken to consist of two parts, each providing an interaction $V_I(x,t)$ and a time derivative $\partial/\partial t$. For the minimally-invasive clock, the interaction $V_I(x,t)$ must be negligibly small. The invasive interaction provides a finite $V_I(x,t)$ which drives transitions in the system. Each environment degree of freedom provides a time variable for the system. However, it is shown explicitly in ref.\cite{JSBthai} that the time for the interacting part can be synchronised to the unique clock time to give a single $\partial/\partial t$ term. Then the generalisation of \eref{Eq:TDHJE4} to
\begin{equation}
\label{Eq:TDHJE5}
 H_{\mathcal{S}}\left(x, \frac{\partial {\mathrm{S}}}{\partial x}\right) + V_I(x,t) + \frac{\partial {\mathrm{S}}}{\partial t} = 0
\end{equation}
is justified. The $V_I(x,t)$ comes from the external interaction and the $\partial/\partial t$ from the clock. However the condition that $\left(\partial {W_{\mathcal{S}}}/\partial R\right)^{2}$ be negligible, where $R$ is now the coordinate of the interacting environment, is still necessary for the validity of this equation. A good example of an interacting environment in classical mechanics is where an external frictional force is modelled by a time-dependent potential acting on the system.

The quantum case is more complicated in that, to provide a time-dependent potential, the interacting environment must be treated classically, a requirement ignored in most text books.
The situation arises when environments which normally should be treated by quantum mechanics have energy so great that they can be  described by classical mechanics to a good approximation. In fact this classical approximation of perturbing potentials on a quantum system
is the \emph{only source of time dependent Hamiltonians in quantum mechanics}. From this point of view the TDSE is always a mixed quantum-classical equation. 

The prime examples of transition from quantum to classical mechanics are to be found in the description  of the impact of a particle or light beam on a quantum system, for simplicity, call it an atom. If the beam energy is low the projectile must be treated fully quantum-mechanically by the TISE of the composite of (beam + atom). Energy is exchanged between beam and atom in entangled states. When the beam energy greatly exceeds atom energies, the beam motion can be treated by Newton's equations  along a classical trajectory. Then one has the limit of a TDSE for the atom alone and there is no entanglement with the beam. The classical beam motion gives rise to a time-dependent potential $V_I(t)$ acting on the atom. When the beam energy is sufficiently large, one may in addition forget the back-reaction on the beam motion via the potential $U_\mathcal{S}(t)$, decoupling the projectile beam motion entirely from the target atom. 

In the same way, when a light beam consists of a few photons its field must be quantized and treated in a TISE with the atom. There is full entanglement and changes in the atom energy are exactly balanced by changes in the field energy. However  in the case of a very intense beam, where absorption of a few photons by the atom does not affect the beam intensity, it can be treated classically and described by a time-dependent field obeying Maxwell equations.

The two examples above are of applied deterministic classical perturbing environments. Of course there is a huge literature on open quantum systems interacting with environments of essentially infinite dimension e.g a bath of oscillators, leading to a stochastic TDSE or equivalent density matrix descriptions.

\subsection{Corrections to time-dependent equations}

The approximation in deriving the TDSE is the neglect of certain second-order derivatives with respect to $R$. For the classical limit, say of a particle beam, as shown above this involves the neglect of $\frac{1}{2M}\frac{\partial^2 {W_{\varepsilon}}}{\partial R^2}$. This requires that the action varies only slowly over atomic dimensions which is the case where the beam energy and hence the momentum is large on an atomic scale. In the extreme approximation of a \emph{fixed} large momentum $P = \frac{\partial {W_{\varepsilon}}}{\partial R}$
then this second derivative is identically zero. Incidentally this would provide the perfect clock - a point particle moving with constant velocity.

The important approximation necessary to derive the TDSE for the system is the neglect of $\frac{\hbar^2}{2M}\frac{\partial^2 {\psi}}{\partial R^2}$ in \eref{Eq:psidef}. In the limit that the environment is treated classically, one can transform to $t(R)$. For simplicity put $t = MR/P  \equiv R/v$ where $v$ is the constant classical velocity. Then one has a correction term to the TDSE of magnitude
\begin{equation}
\label{Eq:TDSEcorr}
 - \frac{\hbar^2}{2M} \frac{\partial^2 \psi}{\partial R^2} =  - \frac{\hbar^2}{2Mv^2} \frac{\partial^2\psi}{\partial t^2}
\end{equation}
to be compared to the retained term
\begin{equation}
-i\hbar\frac{\partial \psi}{\partial t}.
\end{equation}
Clearly, irrespective of the higher power of $\hbar$, the presence of the classical beam energy in the denominator of the neglected term indicates that this is of small magnitude. Nevertheless, this second derivative is the first  ``quantum'' correction to the half-classical TDSE. A similar correction term has been discussed by Arce \cite{Arce}.

In fact, the corrections to the TDSE could readily be traced experimentally. For example, in the ion-atom collision case discussed above, one could begin with, say a proton, of low velocity (a few eV) necessitating quantisation of the beam and use of the TISE for the composite. Then one could increase the energy successively all the way up to a few keV
where the beam can be treated by classical mechanics providing an effective TDSE for the target atom alone.

The classical case is somewhat more straightforward in that both environment and system always obey classical mechanics. The question is; which classical mechanics ?
In the reduction to a TDHJE for the system alone the term $\frac{1}{2M} \left(\frac{\partial {W_{\mathcal{S}}}}{\partial R}\right)^{2}$, which is the classical analogue of the quantum term $-\frac{\hbar^2}{2M}\frac{\partial^2 {\psi}}{\partial R^2}$ has been neglected. Again in the simplest form of environment motion $R = v t $ this gives
\begin{equation}
\frac{1}{2M} \left(\frac{\partial {W_{\mathcal{S}}}}{\partial R}\right)^{2} =  \frac{1}{2Mv^2} \left(\frac{\partial {W_{\mathcal{S}}}}{\partial t}\right)^{2}
\end{equation}
to be compared to the retained term $\frac{\partial {W_{\mathcal{S}}}}{\partial t}$ in \eref{Eq:TDHJE1}.  Again the correction is lower by at least a factor the inverse of the environment kinetic energy.  Also it is often so that $\frac{\partial {W_{\mathcal{S}}}}{\partial t}$ itself is small so that the square is yet smaller. Quite what is the environment energy to be included is questionable.
The correction is difficult to estimate in detail but some idea can be obtained in the limit that $\mathrm{S}_{\mathcal{S}}(x,t) \approx W_{\mathcal{S}}(x) - E_{\mathcal{S}}t$ where
$E_{\mathcal{S}}$ is the system energy. Then \eref{Eq:TIHJsystem} would become
\begin{equation}
\label{Eq:TIHJsystem2}
 H_{\mathcal{S}}\left(x,\frac{\partial{W_{\mathcal{S}}}}{\partial x}\right) = E_{\mathcal{S}}\left( 1 - \frac{ E_{\mathcal{S}}}{2Mv^2}\right)
\end{equation}
that is, the neglected term gives an effective change in the system energy.

In principle in classical mechanics an interaction with the rest of the universe is unavoidable which would make the correction truly negligible, except perhaps for systems of cosmic size or for integral over astronomical time. Nevertheless, since the TDHJE leads to Hamilton's and Newton's equations, there are in principle corrections to these equations, however small.

\subsection{Quantum gravity}
The main subject of this paper is non-relativistic classical and quantum mechanics, although it has been shown \cite{br12} that in the case of a relativistic single particle the transition from time-independent Dirac equation to a time-dependent one can be made exactly as for the Schr\"odinger equation.  There is another field of relativity which parallels closely the transition from time independence to time dependence which has been given here. This is the problem of time in quantum gravity.

The TIHJE and the TISE refer to closed systems. Clearly, for any finite system this is an approximation to the extent that the interaction with the rest of the universe is ignored.
Any system observed by humans is, of course, in principle open and the extent to which it can be viewed as closed depends upon the accuracy of the measurement upon it. 
The only truly closed system devoid of all external interaction is the whole universe. The dynamics of this is described classically by the field equations of general relativity (GR) which of course contain time as a component of the four-vectors of spacetime. Interestingly, the quantum equation which reduces to the classical GR equation contains only the three-metric i.e. is timeless. The Schr\"odinger-like equation describing the composite (quantum gravity + quantum matter fields) is the Wheeler - de Witt  equation (WDE) $H\Psi = 0$. That is, it is the TISE with total energy put equal to zero, as is reasonable to assume for the entire universe. Specifically the WDE in a compact form reads \cite{KIESi}
\begin{equation}
\label{Eq:WdWfull}
\left(-\frac{\hbar^2}{2M} G_{ab}\frac{\delta^2}{\delta h_a\delta h_b} + 2Mc^2\sqrt{h}(2\Lambda - \mathcal{R}) + \mathcal{H}_m\right)\Psi = 0.
\end{equation}
Here $M \equiv c^2/(32\pi G)$, where $c$ is the light velocity and $G$ the gravitational constant, $G_{ab}$ and $h_a$ are coefficients of the deWitt metric and the three-metric respectively, $\Lambda$ is the cosmological constant, $\mathcal{R}$ is the three-dimensional Ricci scalar and $h$ is the determinant of the three-metric.
The matter Hamiltonian $\mathcal{H}_m$ for  a matter field $\phi$ is taken as
\begin{equation}
\label{Eq:Hm}
\mathcal{H}_m = \frac{1}{2}\left(-\frac{\hbar^2}{\sqrt{h}}\frac{\delta^2}{\delta\phi^2} + \sqrt{h}(m^2\phi^2 + U(\phi)) + \sqrt{h}h^{ab}\phi_{,a}\phi_{,b}\right)
\end{equation}
where the matter potential $U$ is arbitrary.

Although these equations contain \emph{functional} derivatives with respect to three-metric and matter field functions (which makes any solution exceedingly difficult to obtain) one sees a striking resemblance to the TISE of quantum mechanics \eref{Eq:full2dTISE} written in the form, with the total energy $E \equiv E_{\varepsilon} + E_{\mathcal{S}}$,
\begin{equation}
\left( -\frac{\hbar^2}{2M}\frac{\partial^2\Psi}{\partial R^2} 
 + [V_{\varepsilon}(R)  - E_{\varepsilon}]+ H_\mathcal{S} \right) \Psi(x,R) = 0
\end{equation}
and
\begin{equation}
H_\mathcal{S} \equiv  -\frac{\hbar^2}{2m}\frac{\partial^2}{\partial x^2} +  [V_{\mathcal{S}} - E_\mathcal{S}] + V_I(x,R).
\end{equation}
Clearly, gravity (the three-metric) is the environment and the matter field is the quantum system. The energy density term $2Mc^2\sqrt{h}(2\Lambda - \mathcal{R})$ of the gravitational field plays the role of the environment energy $V_{\varepsilon}(R)  - E_{\varepsilon}$. Similarly the energy term $\sqrt{h}(m^2\phi^2 + U(\phi))$ of the matter field is equivalent to the energy $V_{\mathcal{S}} - E_\mathcal{S}$ of the quantum system. The interaction minimal coupling between three-metric and matter $ \sqrt{h}h^{ab}\phi_{,a}\phi_{,b}$ is the analogue of the interaction potential  $V_I(x,R)$ which here has been included as part of the system Hamiltonian.

Apart from the timeless approach of Barbour \cite{bar1} , the absence of time in the fundamental equation of quantum gravity has been viewed as a problem \cite{And1}. However, the analogy with non-relativistic quantum mechanics would indicate no conceptual difficulty. A time-independent quantum state of a nucleus for example is a superposition of various states of different character and apportionment of energy between the nucleons, but all at the same total energy. The quantum solution says that the universe exists in various states of different character, products of states of the three-metric and corresponding states of matter, all at the same total energy zero but characterised by 
different apportioning of energy density between metric and matter. The difference with the nucleus is that there is nothing exterior to the universe so no observation can be made. However, a human is a negligibly small part of the universe and so effectively can make a non-invasive observation of the various states of the universe. Then time arises when the matter is observed and is introduced only in the limit that the gravitational environment becomes classical, obeying Einstein's classical field equations. Then one obtains a time-dependent ``Schr\"odinger'' equation for the matter fields. This is absolutely parallel to the classical limit of a massive environment of section VB above.  There the TDSE for the quantum system with an effective time-dependent Hamiltonian arises from a classical environment obeying Newton or Maxwell equations.

There are many papers employing the above strategy (examples are \cite{And1} - \cite{Vile}). Almost all begin with a product BO Ansatz and then make a WKB approximation for the gravity term in the product (so-called semi-classical gravity). Although it is recognised that a more exact form is a linear combination of such products, as in \eref{Eq:BOexp}, there have been no attempts to solve the coupled equations. Non-BO terms have been considered explicitly by  Kiefer and Singh \cite{KIESi} . However, they adopt a different strategy by first writing the composite total wavefunction in terms of a total action and then  proceed by an expansion of this action in powers of $M$. Their result  is equivalent to a functional of BO form  (translating their notation to the notation of this paper)
\begin{equation}
\Psi \approx \exp{\left(\frac{i}{\hbar}MW(h_{ab})\right)}~\psi(\phi,h_{ab}).
\end{equation}
Here the functional $W$ satisfies the classical HJ equation
\begin{equation}
\frac{1}{2}G_{ab}\frac{\delta W}{\delta h_a}\frac{\delta W}{\delta h_b} + V(h_a) = 0
\end{equation}
which is equivalent to Einstein's classical equations. This is the direct analogue of the classical mechanics HJ \eref{Eq:HJenviron1} used to derive the WKB environment wavefunction. Using this result one derives a ``TDSE'' (also called the Tomonaga-Schwinger equation) for
the matter field in a classical spacetime background
\begin{equation}
\mathcal{H}_m \psi = i\hbar G_{ab}\frac{\delta W}{\delta h_a}\frac{\delta \psi}{\delta h_b} \equiv  i \hbar \frac{\delta \psi}{\delta \tau}
\end{equation}
the analogue of \eref{Eq:TDSE1}.
Kiefer and Singh then show that in higher order there are correction terms, as in the case \eref{Eq:TDSEcorr} of the TDSE. However, since their development is as a power series the precise connection to the correction exposed here is not clear.

Several points can be made regarding the implications of the simple problem of  quantum mechanics treated here for the much more difficult problem of quantum gravity.

A. \quad In the applications to quantum gravity the single-product form of the wavefunction $\Psi$ is usually referred to as the BO form. It does not seem to have been appreciated that the single product wavefunction for the universe is, in principle, exact as given by the coupled equations  (\ref{Eq:chidef}) and (\ref{Eq:psidef}) in quantum mechanics.

B.\quad A sum of products of BO wavefunctions has often been suggested for quantum gravity. The simpler strategy of the use of an entangled sum of product wavefunctions of the form \eref{Eq:entang2} has not been employed. This would lead to coupled equations for quantum gravity of the simpler form \eref{Eq:close coupled} where the reaction of matter on the gravitational wavefunctions is explicit and vice versa. Again the WKB approximation would give the time-dependent coupled equations for the matter field alone, as in \eref{Eq:coupltime}. This involves semi-classical gravity interacting via the minimal coupling term of  \eref{Eq:Hm} only (the analogue of $V_I(x,R)$) and the wavefunctions of the different three-metric states will transform into occupation amplitudes for the matter states.

C.\quad There is another problem frequently referred to in connection with quantum gravity. This is the question as to why, in the real TISE \cite{Cramer} or WDE \cite{bar2} \cite{And1}, the wavefunction is complex in general, or equivalently, that the action function in $\Psi = e^{(iW/\hbar)}$ is complex .  Again a consideration from ordinary quantum mechanics sheds some light on this ``problem''.
Firstly, it is clear that the assumption of a complex action for the environment is the origin of the factor $i$ multiplying the time derivative in the TDSE. Since the TISE and the WDE  are real, there seems no compulsion to introduce a complex wavefunction, whose real and imaginary parts both must satisfy the equation. The necessity of this is traced to the description of free rather than bound motion. This point does not appear to have been appreciated hitherto in the various arguments advanced to justify a complex wavefunction. In the points below reference is to the TISE but apply equally well to the WDE of quantum gravity.

The steps leading to a complex time-independent wavefunction are as follows:

1. As already stated, a wavefunction or functional must be an exponential function of the action since, for many particles the action is additive but a wavefunction multiplicative.

2. Bound-state (classically spatially confined) wavefunctions can always be chosen real. The resulting wavefunction must be normalisable i.e. square integrable.

3. Mathematically a solution to the real TISE can be written exactly in the form
\begin{equation}
\Psi = e^{\frac{i}{\hbar}W} 
\end{equation}
where W is a complex function. The splitting off of the factor $i$ in the exponential is a matter of convention, justified below.

4. Momenta are defined as in classical mechanics, for example, $p_R = \partial W/\partial R$.

5. Real wavefunctions correspond either to a purely imaginary $W$ or to the combination $\Psi + \Psi^*$.

6. The necessity of complexity of the wavefunction arises when one or more particles execute what would  classically be \emph{unconfined} directional motion, practically either translation or rotation. Then the action must have a real part to give a wavefunction describing such directed change of position. 
Wavefunctions of this type are complex. The simplest examples for translation and rotation are of the form $\Psi = e^{\frac{i}{\hbar}W}$, where $W$ is the \emph{real classical action}. This is also the reason for the convention of splitting off the factor $i$ in the general case where $W$ is complex. Then the real part of the quantum action corresponds to the real classical action.

7.  Finally, to ensure $p_R = \partial W/\partial R$, the corresponding quantum operator must be $p_R = -i\hbar \partial/\partial R$. This is the origin of the factor $i\hbar$ in the TDSE and the complexity of its solutions.

These points are most simply illustrated by the TISE in one dimension which reduces to 
\begin{equation}
\frac{d^2}{dx^2}\phi + k^2\phi = 0
\end{equation}
with real solutions $sin(kx)$ and $cos(kx)$.  These real solutions give the bound state standing waves in the case of confinement in a box of length $L$. A bound standing wave has no direction.
When the particle becomes free it can be detected in the $+x$ or the $-x$ direction. A directed wave can be formed as a linear combination of the two bound solutions i.e. $\phi(x) = cos (kx) + a~ sin(kx)$ where $a$ is a complex number. Merzbacher \cite{Merz} has shown that if the resulting $x$ dependence is invariant under translation (corresponding to conservation of linear momentum) then $a = \pm i$ i.e. $\phi = e^{\pm ikx}$ and $W = kx = px/\hbar$ the real classical action. The necessity of this form arises simply from our ability to distinguish position change occurring from left to right or vice versa. These solutions of the real TISE are complex. Similar considerations apply to  complex wavefunctions describing rotation $e^{(\pm im\phi)}$ or spherical waves $e^{(\pm ikR)}/R$  giving expansion or contraction. We can detect the direction of shape positional changes.

D. \quad There has been much speculation as to the role of decoherence due to environment interaction on the wavefunction (or density matrix) of quantum gravity \cite{Zeh} \cite{KIESi}. Halliwell \cite{Halli} discusses  ``the manner in which the gravitational field becomes classical in quantum cosmology''. To become classical it is necessary that quantum entanglement disappears, as outlined in this paper. However, whether background interactions are responsible e.g. \cite{Halli} ``that the density matrix of the
Universe will de-cohere if the long-wavelength modes of an inhomogeneous massless scalar field are
traced out.''  remains speculation. There has also been discussion of the effect of decoherence on the expanding and contracting parts of the Hartle-Hawking wavefunction of the universe \cite{HHaw}, the analogues of the expanding and contracting spherical waves $e^{(\pm ikR)}/R$.

 In quantum gravity the consequences of a quantum mechanical result known as the ``imaging theorem'' \cite{Kemble} \cite{JimII} do not appear to have been considered. This theorem shows that any quantum wavefunction propagating to large distances or times  (more accurately to large accumulated values of the action)
shows a behaviour in which position, momentum and time appear in their \emph{classical} relationship. In other words, at macroscopic distances, classical behaviour appears independently of any \emph{external} de-cohering interactions. However, as explained in \cite{JimII}, the phase propagation in time of the wavefunction leads to an effective \emph{internal} de-coherence via a stationary phase approximation.

\section{Conclusions}

The standpoint is adopted that all time occurring in dynamics is relative in that, if a clock is used to measure time, one is quantifying positional changes of an observed object by
comparison with standard positional changes of a generalised clock pointer. A closed composite of several parts is timeless and its states in phase space are described by the classical TIHJE or the quantum TISE. Observation of one part (the system) by another (the environment) is an invasive action requiring interaction. Separating the total action in an ``adiabatic'' form allows an approximate time-dependent dynamics to emerge in which the environment acts as clock. To function as a clock the interaction with the system must be negligible.

The mathematical development is almost identical in classical and quantum cases. However the quantum case is more interesting in that the linearity of the TISE allows more flexibility of solution. Additionally,  in order to achieve a real classical time variable in the quantum case, the environment must itself behave classically so that quantum entanglement of system and environment is destroyed. Strongly interacting parts of the environment give rise to effective time-dependent potentials and in the quantum case this time arises from classical motion of the environment.  The resulting TDHJE and TDSE time-dependent equations for the observed system are approximate and some estimate of the magnitude of the corrections is given.

 Suggestions are made as to how the results of this study may throw light on the parallel problem of time dependence in quantum gravity. It is argued that the complexity of the solution of the  real TISE and the resulting complexity of the TDSE has its root in the description of the \emph{direction} of  unconstrained positional change.

\appendix
\section{The quantum time}
Since the classical time arises in the semi-classical limit of a quantum wavefunction, it is interesting to examine this limit from a general definition of a quantity with dimensions of time.
Hence, first a `quantum time' $\tau$ is defined  w.r.t any wavefunction $\mathcal{K}(R)$ as
\begin{equation}
\label{Eq:def tau}
\tau = \frac{i}{\hbar} M \int^R \frac{\mathcal{K}(R')}{\left(\partial\mathcal{K}/\partial R'\right)}dR'.
\end{equation}
One notes that this `time' is in general a complex quantity and for wholly real wavefunctions is purely imaginary. Such complex times have found application in discussions of tunnelling, for example.

If  $\mathcal{K} = \chi$ is assumed and one substitutes for  $\tau$ in \eref{Eq:psidef}, neglecting the second $R$ derivative gives a new TDSE
\begin{equation}
\label{Eq:TDSEimag}
\left (H_{\mathcal{S}} + V_{\mathcal{I}}(\tau) + U_\mathcal{S}(\tau) -  i\hbar\frac{\partial }{\partial \tau}\right)\psi(x,\tau) = 0
\end{equation}
where the complex $\tau(R)$ is derived from the environment wavefunction.
It would be interesting to explore the consequences of this new equation, intermediate between the full TISE for $\mathcal{C}$ and the real-time TDSE for $\mathcal{S}$ alone.

If one writes  $\mathcal{K} = \chi = A(R) \exp{(\frac{i}{\hbar} \tilde W(R))}$, where $A$ and $\tilde W$ are real functions, one has the complex time
\begin{equation}
\tau  = M \int^R \frac {A(R')}{\left(A \frac{\partial \tilde W}{\partial R'} - i\hbar  \frac{\partial A}{\partial R'}\right)}dR'.
\end{equation}
One notes that even if $\tilde W$ is approximated by the classical action, as in the WKB wavefunction, the time is still complex. However, the classical action is of macroscopic size and hence the term involving $\hbar$ is much smaller and can be neglected. Then the function $A$ cancels to give the real classical time. Also, for potentials which contain powers of $R$ up to the second, it
can be shown that $A$ is independent of $R$ so that again one obtains the real classical time
\begin{equation}
\label{Eq:classtime}
 \tau \equiv t =   M \int^R \frac {dR'}{ \partial \tilde W/\partial R'}  = M \int^R \frac{dR'}{P(R')}.
\end{equation}
 In the extreme case that $V_{\mathcal{E}}$ is also zero one has what is called the `perfect'
 clock.
 This perfect clock is a point particle with fixed linear momentum i.e. moving on a straight line. Then $W = PR$ and  the time is given simply by $t = (MR)/P$. If the classical velocity is introduced as $dR/dt = P/M \equiv v$ then one has the simple classical relation $R = vt$. It is interesting that in this special case, the quantum time \eref{Eq:def tau} and the classical time defined by the semi-classical wavefunction are identical. The exact wavefunction is of course the plane wave
\begin{equation}
\chi(R) = \left(\frac{1}{2\pi\hbar}\right)^{1/2} \exp{\left(\frac{i}{\hbar} PR\right)}.
\end{equation}
However, since the momentum $P$ is fixed, then the spatial wavefunction occupies all space and one requires again the classical limit to define position and so function as a clock.


\begin{thebibliography}{99}
\bibitem{Hasp} M. Haspelmath,  \emph{From Space to Time: Temporal Adverbials in the WorldÕs Languages}. No. 03 in LINCOM Studies in Theoretical Linguistics  (LINCOM Europa, M\"unchen, Newcastle) 1997.
\bibitem{sch1} E. Schr\"{o}dinger, Ann. Phys. {\bf79}, 361, 489 (1926), {\bf81}, 109 (1926).
\bibitem{sch2} E. Schr\"{o}dinger, Ann. Phys.  {\bf81}, 109 (1926).
\bibitem{Isham} C.J. Isham arXiv:gr-qc/9210011 v1 (1992)

\bibitem{And} E. Anderson arXiv:1206.2403v2 (2012)
\bibitem{Cook} David.B.Cook, \emph{Schr\"odinger's Mechanics} (World Scientific, Singapore) 1989, \emph{Probability 
and Schr\"odinger's Mechanics} (World Scientific, Singapore) 2003.
\bibitem{Cramer} J. G .Cramer Revs.Mod.Phys. {\bf 58} 647 {1986}
\bibitem{And1} E. Anderson Class.Quantum Grav. {\bf23} 2491 (2006)
\bibitem{bar2} J.B. Barbour Phys.Rev.D {\bf47} 5422 (1993)


\bibitem{bar1} J.B. Barbour Class.Quantum Grav. {\bf11} 2853 (1994),  {\bf11} 2875 (1994), {\bf20} 1543 (2003), \emph{The End of Time}, (Weidenfeld and Nicholson, London) 1999.

\bibitem{Fine} R. P. Feynman and A. R. Hibbs  \emph{Quantum mechanics and Path Integrals} (McGraw Hill, New
York, 1965), p.84.

\bibitem{lan} C. Lanczos, The Variational Principles of Mechanics (Dover Publications, New York) 1970.
\bibitem{br12} J.S.Briggs and J.M.Rost, Eur.Phys.J. {\bf10}, 311 (2000), Found. of Phys. {\bf31}, 693 (2001).
\bibitem{brma} J.\ S.\ Briggs and J.\ H.\ Macek, Adv. At. Mol. Opt. Phys. {\bf 28} 1 (1991).
\bibitem{JSBthai} J.S.Briggs, S. Boonchui and S. Khemanni, J.Phys. A {\bf 40} 1289 (2007)
\bibitem{Br04} J. S. Briggs in "Nonadiabatic Transition in Quantum Systems",
p. 69, eds. V. I. Osherov and L. I. Ponomarev (Inst. of Problem of Chem. Phys. Chernogolovka) (2004).
\bibitem{Ced} G.Hunter Int.J.Quant.Chem. {\bf9} 237 (1975),\ A.Abedi, N.T. Maitra and E.K.U. Gross Phys.Rev.Letts. {\bf105} 123002 (2010), \ L.S. Cederbaum J.Chem.Phys.    {\bf138} 224110 (2013).
\bibitem{Arce} J.C.Arce  Phys.Rev.A {\bf85} 042108 (2012)

\bibitem{Zeh} H.D. Zeh \emph{The physical basis of the direction of time} (Berlin: Springer) 1989
\bibitem{KIESi} C.Kiefer and T.P. Singh Phys.Rev.D{\bf44} 1067 (1991)
\bibitem{Banksy} T. Banks  Nucl. Phys.  B{\bf249} 332 (1985)
\bibitem{BHW} R. Brout, G. Horwitz and D. Weil Phys.Letts. {\bf192} 318 (1987)
\bibitem{Vile} A.Vilenkin Phys.Rev.D {\bf39} 1116 (1989)
\bibitem{Halli} J.J. Halliwell Phys.Rev.D {\bf39} 2912 (1989)
\bibitem{HHaw} J.B. Hartle and S.W. Hawking Phys.Rev.D {\bf28} 2960 (1983)
\bibitem{Kemble}E. C. Kemble, \emph{Fundamental Principles of Quantum Mechanics with Elementary Applications}, (McGraw Hill, 1937). 
\bibitem{JimII} J.S.Briggs and J.M.Feagin, J. Phys. B: At. Mol. Opt. Phys. {\bf46} (2013) 025202,\\ J.M. Feagin and J.S.Briggs, J. Phys. B: At. Mol. Opt. Phys. {\bf47} 115202 (2014).
\bibitem{Merz} E. Merzbacher \emph{Quantum Mechanics 3rd Edition} ( Wiley: New York) p.14

\end{thebibliography}
\end{document}